\documentclass[aps, prb, twocolumn,floats,showpacs]{revtex4-2}

\usepackage{amsmath,amssymb,bm}
\usepackage{graphicx}
\usepackage{subcaption}
\usepackage{float}
\usepackage{makecell}
\usepackage{pifont}
 \usepackage{ragged2e}
\usepackage[colorlinks=true, letterpaper=true, pdfstartview=FitV, linkcolor=blue, citecolor=blue, urlcolor=blue]{hyperref}

\usepackage{color}

\def\eqa{\begin{eqnarray}}
\def\eea{\end{eqnarray}}
\newcommand{\eq}{\begin{equation}}
\newcommand{\ee}{\end{equation}}

\usepackage{physics}

\begin{document}
\title{Topological constraints on the electronic band structure of hexagonal lattice in a magnetic field}
\author{Qi Gao$^{1}$}
\author{Wei Chen$^{1,2}$} \email{chenweiphy@nju.edu.cn}
\affiliation{$^{1}$National Laboratory of Solid State Microstructures and School of Physics, Nanjing University, Nanjing, China}
\affiliation{$^2$Collaborative Innovation Center of Advanced Microstructures, Nanjing University, Nanjing, China}

\begin{abstract}
The impact of projective lattice symmetry on electronic band structures has attracted significant attention in recent years, particularly in light of growing experimental studies of two-dimensional hexagonal materials in magnetic fields. Yet, most theoretical work to date has focused on the square lattice due to its relative simplicity. In this work, we investigate the role of projective lattice symmetry, particularly the translation, rotational  and sublattice symmetry,  in a hexagonal lattice with rational magnetic flux, emphasizing the resulting topological constraints on the electronic band structure. We show that, at $\pi$ flux, the projective lattice symmetry in the hexagonal lattice enforces novel Dirac band touchings at $E\neq 0$, and for general rational flux it constrains the number of Dirac points at $E=0$. We further analyze the symmetry-imposed constraints on the Chern numbers of both isolated gapped bands and band multiplets connected by Dirac-point touchings. Our results demonstrate that these constraints in the hexagonal lattice differ substantially from those in the square lattice.
\end{abstract}

\maketitle

\section{Introduction}
Symmetry has profoundly shaped our understanding of lattice electronic band structures since the birth of band theory \cite{dresselhaus2007group}.\quad  Symmetry groups not only determine band features and classification, but also protect novel excitations \cite{Symmetry-indicators,Quantitative-mappings,Classification-Review,QSHE,Z2,Z2-1,Z2Inversion}.\quad Moreover, recent breakthrough in topological quantum chemistry  \cite{Topological-quantum-chemistry,TQC-supplement1,TQC-supplement2,TQC-supplement3,TQC-supplement4,TQC-supplement5} provides a practical method for identifying topological phases by analyzing the irreducible representations of the little cogroup at each high-symmetry momentum of a given material \cite{TQC-Search}, placing symmetry in an even more prominent role.

Although it was recognized as early as the 1960s that symmetry analysis of electronic band structures must be substantially revised in the presence of a magnetic field, where lattice symmetry is described by projective rather than ordinary groups \cite{The-Mathematical-Theory,ZakTranslation}, the implications of this modification have attracted broad interest only in recent years \cite{k.p,Mirror-Chern,Xiao2024,Chen2023,Chen2022,Hofstadter-Topology,Hofstadter-Topology-with-Real-Space,Symmetry.indicatorsin.commensurate.magnetic.flux}. 
It has been shown that the projective lattice symmetry leads to novel symmetry protected band touchings, and topological insulators with Mobius edge states in square lattice \cite{Z2-projective}. Moreover, the projective lattice symmetry can also result in topologcial constraints on the Chern number of electronic bands in the magnetic field \cite{ManyChern,Zak}.

Most theoretical studies on the effects of projective lattice symmetry so far have focused on the square lattice due to its simplicity \cite{Z2-projective,Chen2022}. However, the majority real two-dimensional (2D) materials, such as graphene \cite{Graphene}, boron-nitride, and transition-metal dichalcogenides, adopt a hexagonal structure, which is generally the most stable lattice arrangement in two dimensions \cite{Ashcroft76}. The behavior of such 2D hexagonal materials in magnetic fields has attracted growing experimental interest in recent years, which calls for a detailed theoretical study of the hexagonal lattice in such condition.

In this work, we investigate the electronic band structure of a hexagonal lattice in a magnetic field with rational flux $\phi=2\pi p/q$ ($p, q$ are coprime integers) in each unit cell \cite{VHS}, highlighting two aspects of the band structure: one is the projective lattice symmetry enforced Dirac band touchings, and the other is the topological constraints on the Chern number of the electronic bands. We show that the projective lattice symmetry in hexagonal lattice produces band-structure characteristics and topological constraints markedly different from those of the square lattice \cite{Hofstadter1976}.

 The main results we obtained are: (1) The combined time reversal and translation symmetry in the hexagonal lattice with flux $\phi=\pi$ enforces novel Dirac band touchings at $E\neq 0 $ at the high symmetry point $X=(\pm \pi/2, 0)$ in the magnetic Brillouin zone (MBZ). These Dirac band touchings at $E\neq 0$ do not exist in the square lattice Hofstadter model.
 (2)We show that when the hopping parameters in the hexagonal lattice admit zero-energy solutions of the Hamiltonian, 
 the sublattice, translation and ${\mathcal C}_2$ rotation symmetry together result in $2q$ gapless Dirac points at $E=0$ in general. However, at the phase transition points between the gapless and gapped states tuned by the hopping parameters, the $2q$ Dirac points with chirality $\pm 1$ merge to $q$ semi-Dirac points with chirality zero.
(3)While the constraint on the Chern number of the gapped single bands, i.e., $pC_i=1 {\ \rm mod \ } q$, applies for both square and Hexagonal lattice, the constraints on the Chern number of the glued band pair with band touchings, which are still unexplored in previous works, are different for square lattice and hexagonal lattice, as shown in Table \ref{Table_comparison}.
 (4)In the hexagonal lattice, the sublattice symmetry ${\cal S}$
 commutes with translation, whereas in the square lattice (with  $q$ even) it anti-commutes \cite{WEN1989641}. Consequently, breaking primitive translation symmetry yields a topologically trivial insulator in the hexagonal lattice but a nontrivial one in the square lattice \cite{Xiao2024,Dimerized-Hofstadter-Model}.

We note that our theoretical results in this work remain valid for distorted hexagonal lattices, provided that the translation and sublattice symmetries are preserved. Thus, our conclusions also apply to most realistic 2D hexagonal materials with moderate distortions.

 The hexagonal Hofstadter model may be experimentally realized in Moire bilayer superlattice~\cite{Dean2013} and heterostructure~\cite{Geim2013, Hunt2013, Spanton2018}. It can also be simulated by cold atoms in optical lattice~\cite{Bloch2013, Ketterle2013} 
 and photonic crystals~\cite{Kraus2012, Lahini2009}. These systems provide a variety of platforms to testify the results of our work.  

This paper is organized as follows: In Sec.\ref{sec:bands}, we introduce the projective lattice symmetry of the hexagonal lattice in the magnetic field, and then study the Dirac points enforced by the symmetry, including both those at $E\neq 0$ and $E=0$. In Sec.\ref{sec:Chern}, we study the topological constraints on the Chern number imposed by the lattice symmetry, addressing both the isolated single bands and the band multiplets connected by the Dirac points. In Sec.\ref{sec:Compare}, we compare the impacts of the projective lattice symmetry on the electronic structure in the hexagonal lattice and square lattice. In sec.\ref{sec:summary}, we have a discussion and summary of this work.

\section{Projective lattice symmetry protected Dirac points of the hexagonal lattice in a magnetic field}\label{sec:bands}
In this section, we study the projective lattice symmetries of the hexagonal lattice in a magnetic field with rational flux $\phi=2\pi p/q$ where $p, q$ are coprime integers, and the Dirac band touchings of the electronic bands caused by these symmetries.

It has been shown that there exist $2q$ zero-energy Dirac points in the hexagonal lattice with isotropic hoppings and magnetic flux $\phi=2\pi p/q$ \cite{VHS,Stability,2qDirac-Hamilition}.  In this section, we show that other than these zero-energy Dirac points, at flux $\phi=\pi$, the combined symmetry ${\cal T}\circ\mathbf{T}_2$ enforces Dirac band touchings at $E\neq 0$ in the hexagonal lattice, where ${\cal T}$ is the time reversal operator and $\mathbf{T}_2$ is the magnetic translation operator of the primitive lattice (see the following text for definition). These finite energy Dirac points do not exist in the square lattice with magnetic flux $\phi=\pi$ or $Z_2$ gauge field \cite{Hofstadter1976,Z2-projective}. 

For the Dirac points at $E=0$, Ref.\cite{Stability} demonstrated the existence of $2q$ such Dirac points in an ideal hexagonal lattice with isotropic hopping in the magnetic field, protected by the lattice’s sublattice and rotational  symmetries ${\cal C}_2, {\cal C}_3$.
 In this work, we establish the existence of these Dirac points under more general conditions: when the hopping parameters of the lattice fall in the regime where the zero-energy solutions of the  Hamiltonian in the magnetic field are admitted, 
 e.g., $||t_i|^q-|t_j|^q|\leq |t_k|^q\leq ||t_i|^q+|t_j|^q||, i, j, k=1, 2, 3$  for the tight-binding model \cite{VHS}, 
the sublattice, translation and ${\cal C}_2$ rotation symmetry of the hexagonal lattice result in  $2q$ Dirac points at $E=0$ away from the boundary. At the boundary of this regime, the $2q$ Dirac points at $E=0$ merge to $q$ semi-Dirac points at zero energy.

In the remainder of this section, we first introduce the projective symmetries of the hexagonal lattice and then analyze the Dirac band touchings at 
$E\neq 0$ and $E=0$, respectively.

\subsection{Projective lattice symmetry of the hexagonal lattice in a magnetic field}
\label{subsec:symmetry}

\begin{figure}[h]
    \centering \includegraphics[width=0.45\textwidth]{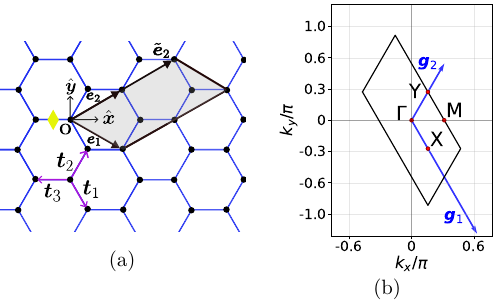}
   \caption{         \justifying
    (a) A real space hexagonal lattice,  where $\mathbf{e}_1, \mathbf{e}_2$ are unit vectors of the primitive lattice. The gray shaded  area is a magnetic unit cell with flux $\phi=\frac{p}{q}\phi_0, \phi_0=h/e$ and magnetic unit vectors $\tilde{\boldsymbol{e}}_{1}
    = \boldsymbol{e} _1, \tilde{\boldsymbol{e}}_{2}=q\boldsymbol{e}_{2}$ ($q=2$ in the figure as an example). The yellow rhombus is the center point of 
    ${\cal C}_2$ rotation.
     (b) The first magnetic Brillouin zone (MBZ) of the Magnetic unit cell, and high symmetry points $\Gamma, X, Y, M$. The reciprocal lattice vectors are $\boldsymbol{g}_1=2\pi(\frac{1}{3}\hat{x}-\frac{1}{\sqrt{3}}\hat{y}),\boldsymbol{g}_2=\frac{2\pi}{q}(\frac{1}{3}\hat{x}+\frac{1}{\sqrt{3}}\hat{y})$. } \label{fig:Lattice} 
\end{figure}
Our setting is a hexagonal lattice  shown in Fig.\ref{fig:Lattice}(a) in a perpendicular magnetic field $\boldsymbol{B}=\partial_x A_y-\partial_y A_x$. The unit vectors of the primitive lattice at zero magnetic field are $\boldsymbol{e}_1=(\frac{3}{2}\hat{x}-\frac{\sqrt{3}}{2}\hat{y})a$ and $\boldsymbol{e}_2=(\frac{3}{2}\hat{x}+ \frac{\sqrt{3}}{2}\hat{y})a$ with $a$ the bond length and $\hat{x}, \hat{y}$ the unit vectors in the $x$ and $y$ direction. Hereafter we set $a=\hbar=e=1$.

In a magnetic field, the Hamiltonian is modified by the Peierls substitution $\boldsymbol{p}=\boldsymbol{P}+\boldsymbol{A}$. To commute with the modified Hamiltonian, 
the symmetry operator $\hat{g}$ in the zero magnetic field must also be modified by a gauge transformation \cite{Hofstadter-Topology,Hofstadter-Topology-with-Real-Space}. Specifically, $\hat{g}$ needs to be replaced by $g\equiv \tilde{G}_g \hat{g} $, where $\tilde{G}_g =e^{i\sum_{\mathbf{r}} \lambda_g(\mathbf{r})c^\dag_{\mathbf{r}} c_\mathbf{r}}$  is a gauge transformation and $\lambda_g(\mathbf{r})$ has been worked out to satisfy \cite{Symmetry.indicatorsin.commensurate.magnetic.flux}
\begin{equation}\label{eq:lambda}
\nabla \lambda_g(\mathbf{r})=\mathbf{A}(\mathbf{r})-R_g \mathbf{A}(g^{-1}\mathbf{r}),
\end{equation}
where $R_g$ is the point group part of $g$. Eq.(\ref{eq:lambda}) determines each $\lambda_g$ up to a constant. In this work we choose the constant such that for translation 
\begin{equation}
\lambda_{\mathbf{T}(\mathbf{a})}(\mathbf{r}) =\int^{\mathbf{r}}_{\mathbf{r}-\mathbf{a}}
\mathbf{A}(\mathbf{r'})\cdot d \mathbf{r'}+\mathbf{B}\cdot (\mathbf{a}\cross \mathbf{r}).
\end{equation}

The general expression for a symmetry operator in a magnetic field is then \cite{Symmetry.indicatorsin.commensurate.magnetic.flux}
\begin{equation}
g=\sum_{\mathbf{r}}e^{i\lambda_g(\hat{g} \mathbf{r})}c^\dag_{\hat{g} \mathbf{r}}c_\mathbf{r}.
\end{equation}

We now consider the translation symmetry in a hexagonal lattice. We denote the translation operators by $\mathbf{e}_1$ and  $\mathbf{e}_2$ as $\mathbf{T}_1$ and $\mathbf{T}_2$ respectively. In this work, we choose the gauge field $\mathbf{A}=\hat{y}(x+\sqrt{3} y)B$ such that for a translation from $\mathbf{r}-\mathbf{e}_i, i=1, 2$ to $\mathbf{r}\equiv m \mathbf{e}_1+n \mathbf{e}_2$, the phase factor 
\begin{eqnarray}
 \lambda_{\mathbf{T}_1}(\mathbf{r})&=&0, \nonumber\\
\lambda_{\mathbf{T}_2}(\mathbf{r})&=&\int^{\mathbf{r}}_{\mathbf{r}-\mathbf{e}_2}
\mathbf{A}(\mathbf{r'})\cdot d \mathbf{r'}+\mathbf{B}\cdot (\mathbf{e}_2\cross \mathbf{r}) \nonumber\\
&=&(-\frac{1}{2}-m+n)\phi. \label{eq:lambda_T2}
\end{eqnarray}

From $\lambda_{\mathbf{T}_1}$ and $\lambda_{\mathbf{T}_2}$, one gets \cite{Symmetry.indicatorsin.commensurate.magnetic.flux}
\begin{equation}\label{eq:non_com}
\mathbf{T}_1 \mathbf{T}_2=e^{i\phi}\mathbf{T}_2 \mathbf{T}_1.
\end{equation}
For rational flux $\phi=B \frac{3\sqrt{3}}{2} a^2=2\pi p/q $ where $p, q$ are coprime integers, $\mathbf{T}_1$ and $\mathbf{T}_2$ do not commute. To make the two translation operators commute with each other, we enlarge the unit cell and choose a $1\cross q$ magnetic unit cell (MUC) with $2q$ sites as shown in Fig.\ref{fig:Lattice}(a). The magnetic unit vectors are $(\boldsymbol{\tilde{e}}_1, \boldsymbol{\tilde{e}}_2)=(\boldsymbol{e}_1, q \boldsymbol{e}_2)$ and the corresponding reciprocal lattice vectors are $\boldsymbol{g}_1=2\pi(\frac{1}{3}\hat{x}-\frac{1}{\sqrt{3}}\hat{y}), 
    \boldsymbol{g}_2=\frac{2\pi}{q}(\frac{1}{3}\hat{x}+\frac{1}{\sqrt{3}}\hat{y})$ as shown in Fig.\ref{fig:Lattice}(b).
    For convenience, we denote the momenta inside the first magnetic Brillouin zone (MBZ) as $\mathbf{k}=k_1 \tilde{\mathbf{g}}_1+k_2 \tilde{\mathbf{g}}_2$ with $\tilde{\mathbf{g}}_1\equiv \mathbf{g}_1/2\pi, \tilde{\mathbf{g}}_2\equiv \mathbf{g}_2/2\pi$ and $k_{1, 2}\in(-\pi, \pi]$. In the following, we will use $\mathbf{k}=(k_1, k_2)$ to label the momentum in the first MBZ.

To distinguish from $\mathbf{T}_1$ and $\mathbf{T}_2$, we denote the magnetic translation operator by $\boldsymbol{\tilde{e}}_1, \boldsymbol{\tilde{e}}_2$ as $\mathbb{T}_1$ and $\mathbb{T}_2$ respectively, which satisfy $\mathbb{T}_1 \mathbb{T}_2=\mathbb{T}_2 \mathbb{T}_1$. The eigenstates of $\mathbb{T}_1, \mathbb{T}_2$ are Bloch states satisfying
\begin{equation}
\mathbb{T}_1|\mathbf{k}\rangle=e^{i k_1} |\mathbf{k}\rangle, \ \ \mathbb{T}_2|\mathbf{k}\rangle=e^{i k_2} |\mathbf{k}\rangle.
\end{equation}
Since $\tilde{\mathbf{e}}=\mathbf{e}$, $\mathbf{T}_1 |\mathbf{k}\rangle=\mathbb{T}_1|\mathbf{k}\rangle$. However, from Eq.(\ref{eq:non_com}), 
\begin{equation}\label{eq:projective_translation}
\mathbf{T}_1 \mathbf{T}_2 |\mathbf{k}\rangle=e^{i(k_1+\phi)} \mathbf{T}_2 |\mathbf{k}\rangle,
\end{equation}
so $\mathbf{T}_2 |\mathbf{k}\rangle$ is an eigenstate of $\mathbf{T}_1$ with eigenvalue $e^{i(k_1+\phi)}$, i.e., the action of $\mathbf{T}_2$ maps a Bloch state at $\mathbf{k}=(k_1, k_2)$ to $\mathbf{k'}=(k_1+\phi, k_2)$.

For the hexagonal lattice with the  $1\cross q$ magnetic unit cell, the ${\cal C}_3, {\cal C}_6$ rotation symmetries  are lost but the ${\cal C}_2$ rotation symmetry remains. The symmetry group of the hexagonal lattice in the magnetic filed then reduces to $p_2$, generated by ${\cal C}_2$ and the magnetic translations $\mathbb{T}_1, \mathbb{T}_2$, for which $\mathbb{T}_M\equiv (\mathbb{T}_1, \mathbb{T}_2)$ is a normal subgroup. The little cogroup, which is the quotient group of the space group mod $\mathbb{T}_M$, is well-defined in the hexagonal lattice.

The time reversal symmetry $\cal{T}$ of the hexagonal lattice with generic magnetic flux $\phi$ is broken except for $\phi=\pi \ \rm mod \ 2\pi$. 

In the next subsections, we will apply the above symmetry properties to analyze the electronic band structure of the hexagonal lattice in a rational magnetic flux $\phi=2\pi p/q$.

\subsection{Electronic bands at rational flux $\phi=2\pi p/q$ and symmetry enforced Dirac points at $E\neq 0$}\label{subsec:touch_band_1}

It has been well-know that symmetries can enforce electronic band touchings \cite{Z2-projective,k.p}. For the hexagonal lattice in the magnetic field, the  Dirac band touchings at $E=0$ have been discovered and partly investigated. However, the symmetry enforced Dirac band touchings at $E\neq 0$ are barely explored.
In this subsection, we study the Dirac band touchings at $E\neq 0$ at the high symmetry points $\pm X$ with momenta $\mathbf{k}=(\pm \pi/2, 0)$ when the flux $\phi=\pi$.

We first show the band structure with Dirac point touchings at different rational flux
obtained from the tight binding Hamiltonian of the hexagonal lattice  in the magnetic field. We then prove that the Dirac point touchings at $\mathbf{k}=(\pm \pi/2, 0)$ at $\phi=\pi$ are due to the combined symmetry $T_2\circ \cal{T}$ at $\pm X$ and exist in a more general Hamiltonian of the hexagonal lattice  than the tight binding Hamiltonian.

The tight binding Hamiltonian of a hexagonal lattice in a magnetic field with nearest neighbor hopping can be written as \cite{VHS}
\begin{equation}\label{eq:real_space}
H=-\sum\limits_{<\boldsymbol{r},\boldsymbol{r}^{\prime}>}t_{\boldsymbol{r},\boldsymbol{r}^{\prime}}
        e^{i\frac{2\pi}{\phi_0}\int_{\boldsymbol{r}}^{\boldsymbol{r}^{\prime}}d\boldsymbol{x}
        \cdot \boldsymbol{A}(\boldsymbol{x})} a_{\boldsymbol{r}}^{\dagger}b_{\boldsymbol{r}^{\prime}}+h.c.
\end{equation}
where $a^\dag_{\mathbf{r}}, b^\dag_{\mathbf{r}}$ are spin-polarized electron creation operators on site $(\mathbf{r}, A)$ and $(\mathbf{r}, B)$ with $\mathbf{r}=m\mathbf{e}_1+n \mathbf{e}_2, m, n\in \mathbb{Z}$, and $t_i, i=1, 2, 3$ are nearest neighbor hopping parameters in Fig.\ref{fig:Lattice}(a). For  $\mathbf{A}=\hat{y}(x+\sqrt{3} y)B$, after working out the Peierls phase in Eq.(\ref{eq:real_space}) and a gauge transformation $a^\dag_{m, n}\rightarrow  e^{i\frac{\pi p}{3q}(m+2n)} a^\dag_{m,n}, b_{m,n}\rightarrow e^{-i \frac{\pi p}{3q}(m+2n)} b_{m, n}$, the tight binding Hamiltonian becomes
\begin{equation}
H=-\sum\limits_{m,n}a_{m,n}^{\dagger}(t_3 b_{m,n}+t_1 \omega^{-n}b_{m+1,n}+t_2 \omega^{n}b_{m,n+1})
+h.c. 
\end{equation}
with  $\omega\equiv e^{2\pi i\frac{p}{q}}$.

We label the field operators in the $1\times q$ MUC as $(a^s_{\mathbf{r}}, b^s_{\mathbf{r}})$ where $\mathbf{r}=\tilde{m}\mathbf{\tilde{e}_1}+ \tilde{n}\mathbf{\tilde{e}_2}, \tilde{m}, \tilde{n}\in \mathbb{Z}$ and $s=0, 1, 2, ... q-1.$ After Fourier transformation to the momentum space, the Hamiltonian can be written as \cite{VHS}
\begin{equation}\label{eq:Hamiltonian_k}
    H=-\sum\limits_{\boldsymbol{k}_{\rm MBZ}}\psi_{\boldsymbol{k}}^{\dagger}
    \mathcal{H}(\boldsymbol{k})\psi_{\boldsymbol{k}},\  
    \mathcal{H}(\boldsymbol{k})= \begin{pmatrix}
        0 & h_{\boldsymbol{k}}\\
        h_{\boldsymbol{k}}^{\dagger} & 0
    \end{pmatrix},\
\end{equation}
where $\psi_{\boldsymbol{k}}=\begin{pmatrix}
        a_{\boldsymbol{k}}^{0 },&
        \cdots,&
        a_{\boldsymbol{k}}^{q-1},&
        b_{\boldsymbol{k}}^{0},&
        \cdots&
        b_{\boldsymbol{k}}^{q-1}&
    \end{pmatrix}^T$ and $h_{\mathbf{k}}$ is a $q\times q$ matrix with non-zero matrix  elements 
\begin{align}
    (h_{\boldsymbol{k}})_{ss}&=t_3+t_1e^{ik_1}\omega^{-s}, s=0,\cdots,q-1,\notag\\
    (h_{\boldsymbol{k}})_{s,s+1}&=t_2\omega^{s},\ s=0,\cdots,q-2, \notag\\
    (h_{\boldsymbol{k}})_{q-1,0}&=t_2\omega^{q-1}e^{ik_2}.
\end{align}
All the other matrix elements of $h_{\mathbf k}$ are zero.

\begin{figure}[H] 
  \centering \includegraphics[width=0.45\textwidth]{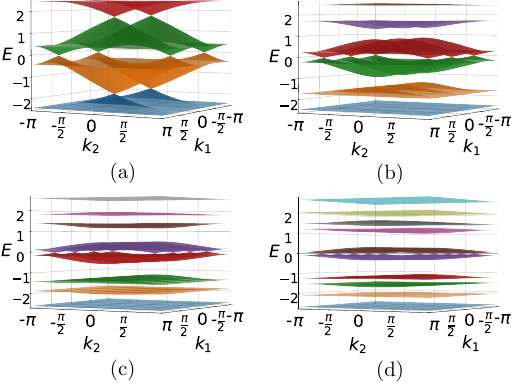}
      \caption{ 
   Electronic bands of the hexagonal Hofstadter model with hopping parameters
   $t_1=t_2=t_3=1$. (a)  $\phi=\pi$  
      (b) $\phi=2\pi/3$   (c) $\phi=2\pi/4$ 
     (d) $\phi=2\pi/5$.}
      \label{fig:3D Band} 
 \end{figure}

We plot the electronic band structure in the first MBZ of the above Hamiltonian with $t_1=t_2=t_3$ for different magnetic fluxes $\phi=\pi, \frac{2\pi}{3}, \frac{2\pi}{4}, \frac{2\pi}{5}$ in Fig.\ref{fig:3D Band}(a), (b), (c), (d) respectively~\cite{Dataset}. All the bands exhibit $2q$ Dirac point touchings at $E=0$, consistent with the results in \cite{Stability,2qDirac-Hamilition} for isotropic hopping.
However, as shown in Fig.\ref{fig:3D Band}(a), for $\phi=\pi$ or $p/q=1/2$, there are Dirac band touchings at $\pm X=(\pm \pi/2, 0)$  and energy $\pm E \neq 0$. We prove in the following that these Dirac band touchings are enforced by the $T_2\circ \cal{T}$ symmetry at $\pm X$ for $\phi=\pi$ and exist for more general Hamiltonians than Eq.(\ref{eq:Hamiltonian_k}) in the hexagonal lattice.

Without loss of generality, we present the proof for the Dirac band touching at $X=(\pi/2, 0)$. Since the time reversal operator $\cal{T}$ maps $X$ to $(-\pi/2, 0)$, and $\mathbf{T}_2$ shifts $(-\pi/2, 0)$ to $(-\pi/2+\phi, 0)=(\pi/2, 0)$ for $\phi=\pi$, $X$ is invariant under the combined operation $\mathbf{T}_2\circ \cal{T}$ for $\phi=\pi$, i.e., $\mathbf{T}_2\circ \cal{T}$ is a symmetry operator at $X$. The Hamiltonian at $X$ should then commute with $\mathbf{T}_2\circ \cal{T}$ for $\phi=\pi$. We can use this condition to construct the general form of the Hamiltonian at $X$ and study the energy spectrum at this point.

From Eq.(\ref{eq:lambda}), for $\mathbf{r}=m \mathbf{e}_1+n \mathbf{e}_2$,
$\lambda_{\mathbf{T}_{2}}(\boldsymbol{r}) =(-\frac{1}{2}+m-n)\pi$ for $\phi=\pi$. 
The $\mathbf{T}_2$ operator in real space in a magnetic field with $\phi=\pi$ can then be written as 
\begin{equation}\label{eq:T_2}
    \mathbf{T}_{2}(\mathbf{r})=\sum\limits_{m,n}e^{i(-\frac{1}{2}+m-n)\pi}
    [a_{m,n}^{\dagger}a_{m,n-1}+b_{m,n}^{\dagger}b_{m,n-1}].
\end{equation}

To Fourier transform $\mathbf{T}_{2}(\mathbf{r})$ to the momentum space, we need to write $\mathbf{T}_{2}(\mathbf{r})$ in terms of field operators in the MUC. For $\phi=\pi$, there are two primitive unit cells, i.e., four atomic sites in a MUC. We label the field operators on the four sites as $(a^0_{\tilde{m},\tilde{n}}, a^1_{\tilde{m},\tilde{n}}, b^0_{\tilde{m},\tilde{n}}, b^1_{\tilde{m},\tilde{n}})$, where $(\tilde{m}, \tilde{n})$ labels position of the MUC, $s=0, 1$ labels the s-th primitive unit cell in the MUC, and $a, b$ are field operators on the $A$ and $B$ sublattice. The translation operator $\mathbf{T}_2$ in Eq.(\ref{eq:T_2}) can then be rewritten as 
\begin{eqnarray}
 \mathbf{T}_{2}(\mathbf{r})&=&\sum\limits_{\tilde{m},\tilde{n}}e^{i(-\frac{1}{2}+\tilde{m})\pi}[e^{i \pi}a^{1 \dag}_{\tilde{m},\tilde{n}}a^0_{\tilde{m},\tilde{n}}
    +a^{0 \dag}_{\tilde{m},\tilde{n}+1}a^1_{\tilde{m},\tilde{n}} \nonumber\\
    &&\ \ \ \ \ \ \ \ \ \ \   +e^{i \pi}
    b^{1 \dag}_{\tilde{m},\tilde{n}}b^0_{\tilde{m},\tilde{n}}
    +b^{0 \dag}_{\tilde{m},\tilde{n}+1}b^1_{\tilde{m},\tilde{n}}].
\end{eqnarray}

With the Fourier transformation
\begin{eqnarray}
    a_{\tilde{m}, \tilde{n}}&=&\frac{1}{\sqrt{N}}\sum\limits_{k_1,k_2}e^{i(k_1\cdot \tilde{m}+k_2\cdot \tilde{n})}
    a_{k_1,k_2}, \\
    b_{\tilde{m}, \tilde{n}}&=&\frac{1}{\sqrt{N}}\sum\limits_{k_1,k_2}e^{i(k_1\cdot \tilde{m}+k_2\cdot \tilde{n})}
    b_{k_1,k_2},
\end{eqnarray}
we get the $\mathbf{T}_2$ operator in momentum space as
\begin{equation}
    \mathbf{T}_{2}(\mathbf{k})=
        \psi^{\pi \dag}_{k_1+\pi, k_2}
    \hat{\mathbf{T}}_{2}(k_1,k_2)\psi^\pi_{k_1, k_2},
\end{equation}
where $\psi^\pi_{k_1, k_2}\equiv (a_{k_1,k_2}^{0}, a_{k_1,k_2}^{1},b_{k_1,k_2}^{0}, b_{k_1,k_2}^{1})$ and  
\begin{equation}
    \hat{\mathbf{T}}_2(k_1,k_2)=e^{-\frac{i\pi}{2}}\begin{pmatrix}
        0 & e^{-ik_2} & 0 & 0\\
        -1 & 0 & 0 & 0\\
        0 & 0 & 0 & e^{-ik_2}\\
        0 & 0 & -1 & 0
    \end{pmatrix}
\end{equation}
 is the projective representation of 
the magnetic translation operator $\mathbf{T}_2$ at generic momentum $(k_1, k_2)$, which satisfy $\hat{\mathbf{T}}_2(k_1,k_2)\hat{\mathbf{T}}_2^{\dagger}(k_1,k_2)=\mathbf{I}$.

At $X$ point, $\mathbf{T}_2\circ {\cal T}$ is a symmetry operator, so it commutes with the Hamiltonian matrix at $X$.
For spinless systems, the time reversal operator acts as a complex conjugate operator, i.e., $\cal{T}=\cal{K}$. The commutation relationship $[\mathbf{T}_2\circ {\cal T}, {\cal H}_X]=0$ at $X$ then reduces to
\begin{equation}\label{eq:H_X_comm}
\hat{\mathbf{T}}_2 (X) {\cal H}_X^*-{\cal H}_X \hat{\mathbf{T}}_2 (X)=0.
\end{equation}

We now construct the general form of the Hamiltonian matrix ${\cal H}_X$ at $X$ satisfying  
Eq.(\ref{eq:H_X_comm}). A general $4*4$ Hermitian Hamiltonian matrix can be expanded by the 16 Dirac matrices:
\begin{align}
    &\Gamma^0 =\mathbf{I}, &
    &\Gamma^1 = \sigma_3 \otimes \sigma_2, &
    &\Gamma^2 = \sigma_2 \otimes \sigma_0, & \nonumber\\ 
    &\Gamma^3 = \sigma_3 \otimes \sigma_1, & 
    &\Gamma^4 = \sigma_1 \otimes \sigma_0, &
    &\Gamma^5 = \sigma_3 \otimes \sigma_3, & \nonumber\\
    &\Gamma^{12} = i\Gamma^1\Gamma^2,   &
    &\Gamma^{13} = i\Gamma^1\Gamma^3,   &
    &\Gamma^{14} = i\Gamma^1\Gamma^4,   & \nonumber\\
    &\Gamma^{15} = i\Gamma^1\Gamma^5,   &
    & \Gamma^{23} = i\Gamma^2\Gamma^3,  &
    &\Gamma^{24} = i\Gamma^2\Gamma^4,   & \nonumber\\
    & \Gamma^{25} = i\Gamma^2\Gamma^5,  &
    &\Gamma^{34} = i\Gamma^3\Gamma^4,   &
    &\Gamma^{35}= i\Gamma^3\Gamma^5,    & \nonumber\\
    &\Gamma^{45} = i\Gamma^4\Gamma^5,
\end{align}
where $\mathbf{I}$ is the $4*4$ unit matrix and $\sigma_0, \sigma_1, \sigma_2, \sigma_3$ are the $2*2$ Pauli matrices.

The Hamiltonian matrix ${\cal H}_X$ can be expanded as 
\begin{align}\label{eq:H_X_expand}
    {\mathcal H}_X &= a_0\mathbf{I} + a_1\Gamma^1 + a_2\Gamma^2 + a_3\Gamma^3 + a_4\Gamma^4 + a_5\Gamma^5 \notag \\
    &\  +  a_{12}\Gamma^{12} + a_{13}\Gamma^{13} + a_{14}\Gamma^{14} + a_{15}\Gamma^{15} + a_{23}\Gamma^{23} \notag \\
    &\   +  a_{24}\Gamma^{24} + a_{25}\Gamma^{25} + a_{34}\Gamma^{34} + a_{35}\Gamma^{35} + a_{45}\Gamma^{45},
\end{align}
with real coefficients $a_0, a_1, ..., a_{45} \in \mathbb{R}$. Applying this form to Eq.(\ref{eq:H_X_comm}) for the $X$ point, i.e., $\mathbf{k}=(\pi/2, 0)$ and solving the coefficients, we get 
\begin{eqnarray}
a_{1}=0,\ a_2=0,\ a_3=0,\ a_5=0,\ a_{12}=0, \nonumber\\
a_{13}=0,\ a_{15}=0,\ a_{23}=0,\ a_{25}=0,\ a_{35}=0.
\end{eqnarray}

The general form of the Hamiltonian matrix ${\cal H}_X$ is then
\begin{eqnarray}\label{eq:H_X_general}
   && \mathcal{H}(X)= \nonumber\\
   && \left(
\begin{array}{cccc}
 a_{0}+a_{24} & 0 & a_{4}-i a_{45} & a_{14}+i a_{34} \\
 0 & a_{0}+a_{24} & -a_{14}+i a_{34} & a_{4}+i a_{45} \\
 a_{4}+i a_{45} & -a_{14}-i a_{34} & a_{0}-a_{24} & 0 \\
 a_{14}-i a_{34} & a_{4}-i a_{45} & 0 & a_{0}-a_{24} \\
\end{array}
\right). \nonumber\\
\end{eqnarray}
This Hamiltonian matrix gives the general form of the energy levels at point $X$, which includes two two-fold degenerate energy levels at finite energy:
\begin{equation}
   E_{\pm}=a_{0}\pm \sqrt{a_{14}^2+a_{24}^2+a_{34}^2+a_{4}^{2}+a_{45}^2}.
\end{equation}
Deviating from $X$, the $\mathbf{T}_2\circ {\cal T}$ symmetry is broken and the two-fold degeneracy at both $E_+$ and $E_-$ is lifted. One then see two isolated Dirac points at $X=(\pi/2, 0)$ with energy $E_+$ and $E_-$ for flux $\phi=\pi$, as shown in Fig.\ref{fig:3D Band}(a). The same proof works for the point $-X=(-\pi/2, 0)$. 

\begin{figure}[h] 
  \centering \includegraphics[width=0.45\textwidth]{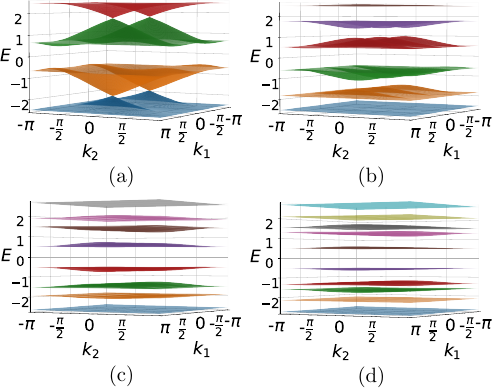}
     \caption{  \justifying
 Electronic bands of the hexagonal lattice with broken sublattice symmetry under A/B site potentials of +0.5/-0.5 and hopping parameters $t_1 = t_2 = t_3 = 1$. (a)  $\phi=\pi$  
      (b) $\phi=2\pi/3$   (c) $\phi=2\pi/4$ 
     (d) $\phi=2\pi/5$.}
      \label{fig:sublattice_broken} 
 \end{figure}

From Eq.(\ref{eq:H_X_general}), one can see that breaking the sublattice symmetry, e.g., adding staggered potential $a_{24}$ and $-a_{24}$ on the $A$ and $B$ sublattice,
does not remove the Dirac band touchings at $\pm X$ for $\phi=\pi$, though it will remove the Dirac band touchings at $E=0$, as shown in Fig.\ref{fig:sublattice_broken}\cite{Dataset}.

\begin{figure}[h] 
  \centering \includegraphics[width=0.45\textwidth]{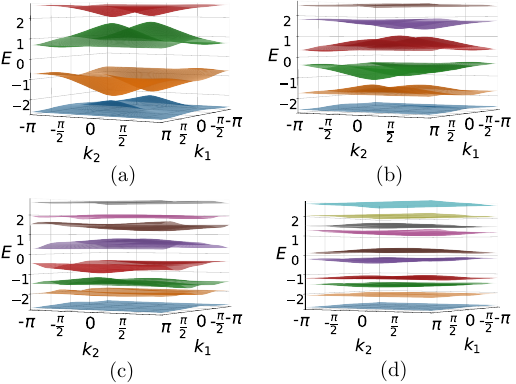}
      \caption{  \justifying
   Energy bands with broken translation symmetry $\mathbf{T}_2$ for hopping parameters $t_1=t_2=t_3=1$: (a) $p/q=1/2$, on-site potential in a MUC on the $A$ and $B$ sublattice $V_A=\{+0.5,+1.0\}$, $V_B=\{-0.5,-1.0\}$
      (b) $p/q=1/3$, on-site potential in a MUC $V_A=\{+0.2,+0.4,+0.6\}$, $V_B=\{-0.2,-0.4,-0.6\}$
     (c) $p/q=1/4$,  on-site potential in a MUC $V_A=\{+0.2,+0.4,+0.6,+0.8\}$, $V_B=\{-0.2,-0.4,-0.6,-0.8\}$
     (d) $p/q=1/5$, on-site potential in a MUC $V_A=\{+0.05,+0.10,+0.15,+0.20,+0.25\}$, $V_B=\{-0.05,-0.10,-0.15,-0.20,-0.25\}$}
      \label{fig:Translation_Broken} 
 \end{figure}

However, if the $\mathbf{T}_2$ translation symmetry of the primitive lattice is broken, e.g, adding different potentials on the two $A$ (or $B$) sites in a magnetic unit cell, the Dirac band touchings at $\pm X$ will be lifted, as shown in Fig.\ref{fig:Translation_Broken}\cite{Dataset}.

We note that though we plot the electronic bands in Fig.\ref{fig:3D Band} with $t_1=t_2=t_3$, the Dirac band touchings at $\pm X$ exist for more general hopping parameters. 
For example, the tight binding Hamiltonian Eq.(\ref{eq:Hamiltonian_k}) 
for generic $t_1, t_2, t_3$ at $\phi=\pi$ takes the form
\begin{eqnarray}\label{eq:TBG_pi_flux}
   && \mathcal{H}(k_1,k_2)= \nonumber\\
    && \begin{pmatrix}
        0 & 0 & t_3+e^{ik_1}t_1 & t_2\\
        0 & 0 & -e^{i k_2}t_2 & t_3-e^{ik_1}t_1\\
        t_3+e^{-ik_1}t_1& -e^{-i k_2}t_2 & 0 & 0\\
        t_2 & t_3-e^{-ik_1}t_1 & 0 & 0
    \end{pmatrix}, \nonumber\\
\end{eqnarray}
which gives two Dirac band touchings at both $X$ and $-X$ with finite energy $E_{\pm}=\pm \sqrt{t_1^2+t_2^2+t_3^2}$ respectively.
The dispersion near the band touching points can be obtained by expanding the eigenenergy near these points, as shown in Appendix A. 
From the expansion, we can see explicitly that the dispersion is linear around the band touching points.

 At other rational magnetic flux $\phi=2\pi p/q$, the combined symmetry $\mathbf{T}_2\circ {\cal T}$
is broken and the symmetry enforced Dirac band touchings at finite energy disappear in the generic case, as shown in Fig.\ref{fig:3D Band}(b)-(d). However, as demonstrated in \cite{VHS}, there may exist hopping-tuned topological phase transitions (TPT) at specific hopping parameters $t_1, t_2, t_3$. At these TPT points, the electronic bands close gap at finite energy with Dirac band touchings. 
Since these Dirac band touchings are not enforced by lattice symmetry and there is no easy and general  solution to the hopping parameters where the TPTs occur, we will not explore these Dirac band touchings further in this work. More discussion about the properties of these TPT points can be found in \cite{VHS}.

\subsection{Band touchings at $E=0$ of the hexagonal Hofstadter model at flux $\phi=2\pi p/q$}\label{subsec:touch_band_2}
The band touchings at $E=0$ of the hexagonal Hofstadter model have been studied partly in \cite{VHS,Stability,2qDirac-Hamilition}
In Ref.\cite{VHS}, the authors solved the bands of this system from the tight binding Hamiltonian and discovered the existence of band touching points at zero energy for hopping parameters  satisfying
\begin{equation}\label{eq:Zero_energy_condition}
||t_i|^q-|t_j|^q|\leq |t_k|^q\leq ||t_i|^q+|t_j|^q||, 
\end{equation}
where $i, j, k$ are any of the three distinct values of $1, 2, 3$.   
Ref.\cite{Stability} demonstrated the existence and stability of $2q$ band touching points at zero energy of an ideal hexagonal lattice with isotropic hopping in the magnetic field protected by the sublattice and ${\cal C}_2, {\cal C}_3$ rotation symmetry.

In this subsection, we supplement the above study by showing that 
when the hopping parameters fall in the regime of Eq.(\ref{eq:Zero_energy_condition}), 
i.e., the zero energy solutions are admitted, the translation and sublattice symmetry of the hexagonal lattice guarantee the existence of $q$ touching points at $E=0$. For the hopping parameters away from the boundary of Eq.(\ref{eq:Zero_energy_condition}), the ${\cal C}_2$ rotation symmetry further doubles the number of band touching points at $E=0$ and results in $2q$ touching points. 
For the hopping parameters at the boundary of Eq.(\ref{eq:Zero_energy_condition}), the $2q$ band touching points merge to $q$ band touching points, which correspond to the critical states between the gapless and gapped phases at $E=0$.

Before the proof of the above results, we have a brief study of the sublattice symmetry of the hexagonal lattice,
which can be defined as  $\mathcal{S}=
      \tau_z  \otimes  I_{q}$, where $I_{q}$ denotes a $q\times q$ identity matrix acting on the $q$ unit cell of the MUC and $\sigma_z$ is the Pauli matrix acting on the $AB$ sublattice in a unit cell. This operator
anticommutes with the Hamiltonian of the form Eq.(\ref{eq:Hamiltonian_k}), i.e., $\{\mathcal{H}(\boldsymbol{k}),\mathcal{S}\}=0,$
indicating the presence of sublattice symmetry in the system. 
Since the translation in the hexagonal lattice does not exchange the $AB$ sublattice, ${\cal S}$ commutes with the translation operators, i.e.,
$[{\cal S},\mathbf{T}_{1,2}]=0$.

The sublattice symmetry
establishes a relationship between states with positive and negative energies. 
For a state with energy $+E$ described by eigenvalue equation
   \begin{equation}
        \begin{pmatrix}
            0 & h_{\boldsymbol{k}}\\
            h_{\boldsymbol{k}}^{\dagger} & 0
        \end{pmatrix}\begin{pmatrix}
            u_{\mathbf{k}}\\
            v_{\mathbf{k}}
        \end{pmatrix}=+E \begin{pmatrix}
            u_{\mathbf{k}}\\
            v_{\mathbf{k}}
        \end{pmatrix},\ 
    \end{equation}
we get
    \begin{equation}
\begin{pmatrix}
            0 & h_{\boldsymbol{k}}\\
            h_{\boldsymbol{k}}^{\dagger} & 0
        \end{pmatrix}\mathcal{S}\begin{pmatrix}
            u_{\mathbf{k}}\\
            v_{\mathbf{k}}
        \end{pmatrix}=-E\cdot \mathcal{S}\begin{pmatrix}
            u_{\mathbf{k}}\\
            v_{\mathbf{k}}
        \end{pmatrix}.
    \end{equation}

This indicates that if $|\psi_\mathbf{k}\rangle=\begin{pmatrix} 
    u_{\mathbf{k}} \\
    v_{\mathbf{k}}
\end{pmatrix}$ is an eigenstate with energy $+E$, then $\mathcal{S}\begin{pmatrix}
            u_{\mathbf{k}}\\
            v_{\mathbf{k}}
        \end{pmatrix}=\begin{pmatrix}
    u_{\mathbf{k}}\\
    -v_{\mathbf{k}}
\end{pmatrix}$ is an eigenstate with energy $-E$. 
Since ${\cal S}$ commutes with the translation operator, $\mathcal{S}|\psi_{\mathbf{k}}\rangle$ has the same momentum as the state $|\psi_\mathbf{k}\rangle$ but with opposite energy, i.e., the sublattice symmetry connects two bands $n, \bar{n}$ with 
        \begin{equation}
E_n(\mathbf{k})=-E_{\bar{n}}(\mathbf{k}).
        \end{equation}
For each $E=0$ state, the sublattice symmetry results in another degenerate zero-energy state at the same momentum. Deviating from this point, the two states have different energy. The two bands then touch at each zero energy state. For the tight binding model satisfying Eq.(\ref{eq:Zero_energy_condition}), away from the boundary of this equation, we show in Appendix B that the dispersion around the touching point at $E=0$ is linear so it is a Dirac point.

We assume that a zero-energy Dirac point exists at $\mathbf{K}_{0}=(k_{1}^{0},k_{2}^{0})$ for the hexagonal Hofstadter model.
Denote the zero energy state at $\mathbf{K}_0$ as $|\mathbf{K}_0\rangle$. 
Since the magnetic translation operator $\hat{\mathbf{T}}_{2}$ satisfies
\begin{equation}
  \hat{\mathbf{T}}_{2}\mathcal{H}(k_1,k_2)\hat{\mathbf{T}}_{2}^{-1}=\mathcal{H}(k_1+2\pi \frac{p}{q},k_2), 
\end{equation}
$\hat{\mathbf{T}}_{2}|\mathbf{K}_0\rangle$ is also an eigenstate with $E=0$ and momentum 
$\mathbf{K}_1=(k^0_1+2\pi \frac{p}{q}, k^0_2)$. By iteration, $\hat{\mathbf{T}}_{2}$ can generate $q$ distinct zero energy states with different momenta
\begin{equation}\label{eq:momenta_dis}
      \boldsymbol{K}_{j}=(k_{1}^{0}+2\pi j\frac{p}{q},k_{2}^{0})\mod (2\pi,2\pi),\  j=0,1,\cdots,q-1.
\end{equation}
Each zero energy state corresponds to a Dirac point at $E=0$ due to the sublattice symmetry so the sublattice and translation symmetry together result in $q$ Dirac points
at $E=0$.

\begin{figure}[h] 
  \centering \includegraphics[width=0.46\textwidth]{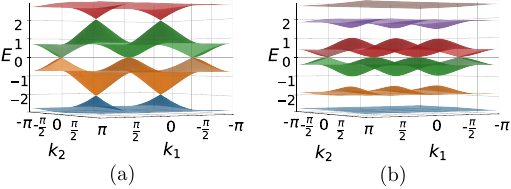}
\caption{ \justifying
Electronic bands of the hexagonal Hofstadter model with $q$ semi-Dirac  points at $E=0$ in the first MBZ. (a) $\phi=\pi$, $t_1=2^{1/2}$, $t_2=t_3=1$, two semi-Dirac points at $E=0$ and  $\mathbf{k}=(0,0)$ and $(\pi,0)$. 
(b) $\phi=2\pi/3$, $t_1=2^{1/3}$, $t_2=t_3=1$, three semi-Dirac  points at $E=0$ and $\mathbf{k}=(-\pi/3,0)$, $(\pi/3,0)$ and $(\pi,0)$. 
}
\label{fig:q_Dirac_points} 
\end{figure}

If one considers further the ${\cal C}_2$ rotation symmetry of the hexagonal lattice, the number of Dirac points at $E=0$ may double. Since the ${\cal C}_2$ symmetry maps the momentum $\mathbf{K}_j$ to $-\mathbf{K}_j$, if the $q$ points $-\mathbf{K}_j, j=0, 1,...,q$ do not coincide with $\mathbf{K}_j, j=0, 1, ..., q$, one gets extra $q$ Dirac points at $E=0$ and there are totally $2q$ Dirac points at $E=0$. The isotropic hopping case with $t_1=t_2=t_3$ belongs to this type.

If the $q$ momenta $\mathbf{K}_j, j=0, 1, ..., q-1$ in Eq.(\ref{eq:momenta_dis}) coincide with $-\mathbf{K}_j, j=0, 1, ..., q-1$, there are totally $q$ band touching  points at $E=0$. An example of this type is $t_1=2^{1/q}e^{iq\pi}, t_2=t_3=1$ for $\phi=2\pi p/q$, as shown in Fig.\ref{fig:q_Dirac_points}\cite{Dataset}.
From the tight binding Hamiltonian, we can further work out the condition of the existence of  only $q$ band touching points at $E=0$ in the hexagonal lattice. The zero energy solutions of the tight binding Hamiltonian satisfy \cite{VHS}
\begin{equation}\label{eq:zero_energy}
    | t_1^q e^{iq k_{1}-i\pi (q-1)}+t_2^{q}e^{ik_2}+t_3^q |=0. 
\end{equation}
From the above analysis, if there exist only $q$ band touching  points at $E=0$, the $q$ momenta $\mathbf{K}_j, j=0, 1, ..., q-1$ in Eq.(\ref{eq:momenta_dis}) should be symmetric with respect to the origin. For the reason, $k_2=0 \ {\rm or} \ \pi$. If one of the $q$ solutions has $k_1=0$, from Eq.(\ref{eq:zero_energy}) we get the hopping parameters satisfying 
\begin{equation}\label{eq:solution_1}
t^q_3-(-t_1)^q=\pm t_2^q, 
\end{equation}
and the $q$ band touching  points locate at $(2\pi j \frac{p}{q}, 0)\rm \ mod\ (2\pi, 2\pi)$ or $(2\pi j \frac{p}{q}, \pi)\rm \ mod\ (2\pi, 2\pi), j=0, 1, ..., q-1$.

If none of the $q$ solutions has $k_1=0$, there must be a solution with $k_1=\pi/q$ to have only $q$ solutions. From Eq.(\ref{eq:zero_energy}), we get the hopping parameters satisfying 
\begin{equation}\label{eq:solution_2}
(-t_1)^q+t_3^q=\pm t_2^q
\end{equation}
and the  $q$ band touching  points locate at $(\frac{\pi}{q}+ 2\pi j \frac{p}{q}, 0)\rm \ mod\ (2\pi, 2\pi)$ or $(\frac{\pi}{q}+ 2\pi j \frac{p}{q}, \pi)\rm \ mod\ (2\pi, 2\pi), j=0, 1, ..., q-1$.

The locations of the $q$ band touching points  in Fig.\ref{fig:q_Dirac_points} are consistent with the above solutions.
The expansion of the energy band  near one of these band touching points at $\phi=\pi$  in Appendix B shows that the dispersion is parobolic along some direction but linear in the other directions, the band touching points in this case are then semi-Dirac points.

Equation (\ref{eq:solution_1}) and (\ref{eq:solution_2}) correspond to the boundary of Eq.(\ref{eq:Zero_energy_condition}) and  the phase transition points from the gapless to gapped phase at $E=0$ tuned by hopping parameters. At these critical points, the $q$ pairs of Dirac points with chirality $+1$ and $-1$ merge to $q$ semi-Dirac points with chirality zero, which are then unstable and a small perturbation of the hopping parameters can open up a gap. The band structure at the semi-Dirac points becomes parobolic after the gap opening.

To lift the Dirac point degeneracy at $E=0$ in the hexagonal lattice away from the boundary of Eq.(\ref{eq:Zero_energy_condition}), one needs to break the sublattice symmetry, as shown in Fig.\ref{fig:sublattice_broken}. Breaking only the $\mathbf{T}_2$ symmetry does not completely lift the Dirac point degeneracy at $E=0$.

\section{Constraints on the Chern number of the electronic bands of the hexagonal lattice in the magnetic field}\label{sec:Chern}

 Due to the existence of the hopping tuned TPTs in the hexagonal Hofstadter model, the Chern numbers of the electronic bands in this system depend on the hopping parameters. However, they satisfy some general topological constraints imposed by the projective lattice symmetry. This is what we 
 study in this section.
 We then compare these topological constraints in the hexagonal lattice with those in the square lattice and show the difference in the two lattices.

\subsection{Topological constraints on the Chern numbers of the degeneracy lifted bands}\label{subsec:Chern_1}

We first study the case when the Dirac band touchings in the hexagonal lattice are lifted by the symmetry breakings  discussed in the last section. We ignore the accidental Dirac band touching cases at the TPT points at $E\neq 0$ temporarily. Since the $\phi=\pi$ case is different from other flux cases, we discuss this case separately.

\subsubsection{\it The $\phi=\pi$ case}
To lift all the  band touchings at the regime 
$||t_i|^q-|t_j|^q|\leq |t_k|^q\leq ||t_i|^q+|t_j|^q||$, we need to break both the $AB$ sublattice symmetry and the $\mathbf{T}_2$ translation symmetry of the primitive lattice.  
Beyond the regime $||t_i|^q-|t_j|^q|\leq |t_k|^q\leq ||t_i|^q+|t_j|^q||$, we only need to break the $\mathbf{T}_2$ translation symmetry to gap the bands at $E\neq 0$. The magnetic translation symmetry $\mathbb{T}_M$ are still preserved  in both situations.

At $\phi=\pi$, the time reversal symmetry of the system is preserved. For this reason, the Chern number $C$ of each gapped band is zero. Moreover, since the sublattice symmetry operator ${\cal S}$ commutes with the translation operator $\mathbf{T}$ in the hexagonal lattice, the gapped state at $\phi=\pi$ by breaking the $\mathbf{T}_2$
translation symmetry 
is topologically trivial. This is in contrast to the square lattice with $\pi$ flux or the $Z_2$ gauge field in Ref.\cite{Xiao2024,Z2-projective,Dimerized-Hofstadter-Model}, for which the gapped state by breaking one of the translation symmetries is topologically nontrivial although the Chern nubmer is zero because in such systems the sublattice 
${\cal S}$ and translation operator $\mathbf{T}$ anti-commute with each other.

\subsubsection{\it The $\phi\neq \pi$ case}
At $\phi=2\pi p/q$ but not equal to $\pi$, there is no generic symmetry enforced Dirac band touchings at $E\neq 0$, and the  band touchings at $E=0$ occur only when $||t_i|^q-|t_j|^q|\leq |t_k|^q\leq ||t_i|^q+|t_j|^q||$. To lift the band degeneracy in this regime, one needs to break the $AB$ sublattice symmetry. Beyond this regime, the electronic bands are generally gapped except at some TPT points tuned by hopping parameters. The degeneracy lifted electronic bands generally have non-zero Chern number which we study in the following.

{\it Constraint due to translation symmetry.} With all the  band touchings lifted, the hexagonal lattice system exhibits $2q$ single electronic bands for the rational flux $\phi=2\pi p/q$ since each $1\times q$ MUC contains $2q$ atoms.
For systems with the preserved primitive translation symmetry, the Chern number $C$ in the magnetic field satisfies \cite{ManyChern}
\begin{equation}\label{eq:Chern_constraint_1}
e^{2\pi i(\frac{p}{q}C-\bar{\rho})}=1,
\end{equation}
where $\bar{\rho}$ is the average number of the electrons in the original unit cell. Eq.(\ref{eq:Chern_constraint_1}) applies both for the total Chern number with $\bar{\rho}$ being the total filling  and the Chern number of a single band with  $\bar{\rho}$ the single band filling.
Assuming one electron on each lattice site, the total filling factor is then $\bar{\rho}=2$. This corresponds to $2q$ full-filled bands. For each  full-filled single band, the average filling factor $\bar{\rho}=1/q$. The Chern number $C_i$ of each single band then satisfies \cite{Zak,ManyChern}.
\begin{equation}\label{Chern_constraint_2}
pC_i=1 \ {\rm mod} \ q.
\end{equation}
This constraint applies to both hexagonal and square lattice.

Eq.(\ref{Chern_constraint_2}) indicates that there are no single electronic bands
at $\phi=\pi$ if the translation symmetry of the primitive lattice is preserved.  If there were single bands at $\phi=\pi$, the Chern number of each single band must be zero due to the time reversal symmetry, which contradicts with Eq.(\ref{Chern_constraint_2}). This analysis is consistent with the Dirac band touchings at $E\neq 0$ at $\phi=\pi$ we obtained in the last section. Only when the translation symmetry of the primitive lattice is broken may the single bands exist for $\phi=\pi$.

{\it Constraint due to the sublattice symmetry.} For gapped system with sublattice symmetry, e.g., beyond the regime $||t_i|^q-|t_j|^q|\leq |t_k|^q\leq ||t_i|^q+|t_j|^q||$, the sublattice symmetry results in an extra constraint on the Chern number of the bands, i.e., the two bands with energy $E_\mathbf{k}$ and $-E_\mathbf{k}$ related by the sublattice symmetry have the same Chern number.

 We label the $2q$ bands of the hexagonal lattice as $\mu=1, 2, ..., 2q$ in this work.
Consider two single bands with energy $\pm E_\mathbf{k}$ related by the sublattice symmetry. The corresponding eigenstates are respectively
\begin{equation}
    \ket{\psi^{\mu}(\boldsymbol{k})}=\begin{pmatrix}
        u^{\mu}(\boldsymbol{k})\\
        v^{\mu}(\boldsymbol{k})
    \end{pmatrix},\ \ket{\psi^{2q-\mu+1}(\boldsymbol{k})}=\begin{pmatrix}
        u^{\mu}(\boldsymbol{k})\\
        -v^{\mu}(\boldsymbol{k})
    \end{pmatrix}.
\end{equation}
The Berry curvature of the $\mu$-th band with energy $E_\mathbf{k}$ is then \cite{Berry1984}
\begin{align}\label{eq:BC_single_band}
   & \mathcal{F}_{xy}^{\mu}(\boldsymbol{k}) \nonumber\\
   & = i\left[
\bra{ \partial_{k_x} \psi^{\mu}(\boldsymbol{k})}\ket{
\partial_{k_y} \psi^{\mu}(\boldsymbol{k})}
-\bra{ \partial_{k_y} \psi^{\mu}(\boldsymbol{k})}\ket{
\partial_{k_x} \psi^{\mu}(\boldsymbol{k})}
    \right] \nonumber\\
    &=i\bigg(
     \partial_{k_x} [u^{\mu}(\boldsymbol{k})]^{*}\cdot \partial_{k_y}u^{\mu}(\boldsymbol{k})
    +\partial_{k_x}[v^{\mu}(\boldsymbol{k})]^{*}\cdot \partial_{k_y}v^{\mu}(\boldsymbol{k}) \nonumber\\
    &\ \ -\partial_{k_y} [u^{\mu}(\boldsymbol{k})]^{*}\cdot \partial_{k_x}u^{\mu}(\boldsymbol{k})
    -\partial_{k_y} [v^{\mu}(\boldsymbol{k})]^{*}\cdot \partial_{k_x}v^{\mu}(\boldsymbol{k})\bigg).
\end{align}
To compute the Berry curvature $\mathcal{F}_{xy}^{2q-\mu+1}(\boldsymbol{k})$ of the $(2q-\mu+1)$-th band with energy $-E_{\mathbf{k}}$, one only needs to replace $v^\mu(\mathbf{k})$ by 
$-v^\mu(\mathbf{k})$ in the above equation. We can easily see that 
\begin{equation}
\mathcal{F}_{xy}^{\mu}(\boldsymbol{k})=\mathcal{F}_{xy}^{2q-\mu+1}(\boldsymbol{k}), 
\end{equation}
which gives
\begin{equation}\label{eq:C_constraint_ph}
C_\mu=C_{2q-\mu+1}.
\end{equation}

\subsection{Topological constraints on the Chern number of the electronic bands with  band touchings}\label{subsec:Chern_2}
In this subsection, we study the Chern numbers of the electronic bands with Dirac band touchings in the hexagonal lattice. We show that the symmetry enforcing the Dirac band touchings leads to extra constraints on the Chern number of the degenerate bands. The constraints we obtained also apply to the bands with semi-Dirac points. For brevity, we will no longer emphasize the latter case separately in the following.

For the electronic bands with Dirac band touchings, the Chern number of a single band is not well-defined as shown below. However, the Chern number of the pair of  bands with band touchings can still be well-defined and measured by the quantum Hall effect of the system.

We define the  Chern number of such glued band pair adopting the formulation in \cite{Degenerate-Multiplets}. In general one can glue the $M$ bands with band touchings together and define 
the non-Abelian Berry connection of the glued band as \cite{Degenerate-Multiplets, Vanderbilt_2018}
\begin{equation}\label{eq:Berry_connection}
 A_{i}^{mn}(\boldsymbol{k})=
i\bra{\psi^{m}(\boldsymbol{k})}\partial_{k_i}
\ket{\psi^{n}(\boldsymbol{k})}, i=x, y,
\end{equation}
where $m, n=1, ..., M$, and $\ket{\psi^{m}(\boldsymbol{k})}$ is the eigenstate of the $m$-th band.
 Away from the touching points, $\ket{\psi^{m}(\boldsymbol{k})}, m=1, ..., M$ are orthogonal to each other automatically.
 At the touching points, to get a well-defined $A_{i}^{mn}(\boldsymbol{k})$,
 one needs to reorganize the $M$ degenerate eigenstates to obtain $M$ orthogonal eigenstates $\ket{\psi^{m}(\boldsymbol{k})}$.

 The non-Abelian Berry curvature of the glued band is then \cite{Degenerate-Multiplets, Vanderbilt_2018}
\begin{equation}
\mathcal{F}^{mn}_{ij} 
    =\partial_{i}A^{mn}_{j}-\partial_{j}A^{mn}_{i}-i[A^{mn}_{i},A^{mn}_{j}], m, n=1, ..., M
\end{equation}
and the Chern number of the  glued band multiplet is
\begin{align} \label{eq:C_M}
    \mathbb{C}_M&=\frac{1}{2\pi i}\int_{\mathrm{BZ}}
     \mathrm{Tr}[\hat{\mathcal{F}}_{xy}(\boldsymbol{k})]\ dk_xdk_y.
\end{align}
We note that both $A_{i}^{mn}(\boldsymbol{k})$ and  $\mathcal{F}^{mn}_{ij}$ defined above are gauge dependent, but the Chern number $\mathbb{C}_M$ is gauge independent \cite{Degenerate-Multiplets}.

For the system with totally $N$ bands, $\mathbb{C}_M$ can be expanded as 
\begin{eqnarray}\label{eq:C_M_expand}
 \mathbb{C}_M =\frac{1}{2\pi}\int d k_x d k_y \ {\rm Im}\sum_{m=1}^{M}\sum_{l=M+1}^{N} 
 \frac{A^{ml}_yA^{lm}_x-A^{ml}_x A_y^{lm}}{(E_m-E_l)^2}, \nonumber\\
\end{eqnarray}
where $m=1, ..., M$ correspond to the $M$ bands with  band touchings, $l=M+1, ..., N$ correspond to the remaining $N-M$ bands and $A_{i}^{ml}$ has the same definition as in Eq.(\ref{eq:Berry_connection}). For simplicity, we have assumed that the $N-M$ bands are all gapped single bands.
When $M=1$, Eq.(\ref{eq:C_M_expand}) reduces to the Chern number of a single gapped band \cite{Niu_review}.

From Eq.(\ref{eq:C_M_expand}), one can 
check that the sum rule of the Chern number enforced by the completeness of the Hilbert space of all eigenstates is still valid even when there are degenerate bands, i.e., 
\begin{equation}\label{eq:sum_rule}
\sum_i C_i=0,
\end{equation}
where $C_i$ includes the Chern number of the glued band multiplet and the remaining single bands.
 We can then compute the Chern number of the glued band directly from Eq.(\ref{eq:C_M_expand}) or from Eq.(\ref{eq:sum_rule}) by computing the Chern number of all the other single bands.

\begin{table*}[t]
\captionsetup{justification=raggedright,singlelinecheck=off}
\centering 
\caption{Topological constraints on the Chern number of  the hexagonal and square lattice with rational flux  
$\phi=2\pi p/q \neq \pi$. The block with $\cross$ indicates that the corresponding case in the first column does not exist. }\label{Table_comparison} 
\begin{tabular}{|c|c|c|c|}
 \hline
 Hofstadter Model& Hexagonal Lattice &  \multicolumn{2}{ c|}{ Square Lattice}  \\\hline
Total number of bands & $2q $ &  \multicolumn{2}{ c|}{  $q$} \\ \hline 
$E=0$ band touching  points  & $\big\vert \vert t_i\vert^q -\vert t_j\vert^q \big\vert\le \vert t_k\vert^q\le \big\vert 
\vert t_i\vert^q +\vert t_j\vert^q \big\vert$ &  \makecell{$q\in \{\text{Odd\ Number}\}$\\  \ding{53} } &
\makecell{ $q\in \{\text{Even\ Number}\}$\\ $\checkmark$} \\\hline
Constraints for the single band& $pC_i=1\mod q$ & \multicolumn{2}{ c|}{$pC_i=1\mod q$}\\\hline
 \makecell{Constraints for the glued band pair\\ with Dirac points at $E=0$ } & $p\mathbb{C}_2=2\mod 2q$ & \makecell{$q\in \{\text{Odd\ Number}\}$\\  \ding{53} } &
\makecell{ $q\in \{\text{Even\ Number}\}$\\ $ p\mathbb{C}_2=2-q\mod 2q.$}\\\hline
 \makecell{Constraints for the glued band pair\\  with Dirac points at $E\neq 0$}  & $p \mathbb{C}'_2=2\mod q$ & \multicolumn{2}{ c|}{ \ding{53} }\\\hline
Constraint by ${\cal S}$ symmetry &$C_{\mu}=C_{2q-\mu+1}$ & \multicolumn{2}{ c|}{
$C_{\mu}=C_{q-\mu+1}$}\\ \hline
\end{tabular}
\end{table*}

For the system with sublattice symmetry and accidental band touchings at $E\neq 0$, e.g., the TPT points,  
the Chern number of the glued Diarc bands at $E>0$ and $E<0$ related by the sublattice symmetry are also the same.
Indeed, if the touching points are at the $\nu,(\nu+1)$-th bands and the $(2q-\nu),(2q-\nu+1)$-th bands, the Chern numbers of the glued lower and upper band pairs are respectively
\begin{equation}
     C_{-}=\frac{1}{2\pi i}\int_{\mathrm{BZ}}\left[
     \mathcal{F}_{xy}^{\nu}(\boldsymbol{k}) + \mathcal{F}_{xy}^{\nu+1}(\boldsymbol{k})\right]dk_x dk_y,
\end{equation}
\begin{equation}
     C_{+}=\frac{1}{2\pi i}\int_{\mathrm{BZ}}\left[
     \mathcal{F}_{xy}^{2q-\nu}(\boldsymbol{k}) + \mathcal{F}_{xy}^{2q-\nu+1}(\boldsymbol{k})\right]dk_x dk_y,
\end{equation}
where $\mathcal{F}_{xy}^{\nu}$ has the same definition as in Eq.(\ref{eq:BC_single_band}).
Note that both $\mathcal{F}_{xy}^{\nu}(\mathbf{k})$ and $\mathcal{F}_{xy}^{\nu+1}(\mathbf{k})$ depend on the gauge and diverge at the band touching points, but their divergences cancel each other and the sum of them is finite and gauge independent \cite{Berry1984}. The same is true for $\mathcal{F}_{xy}^{2q-\nu}(\boldsymbol{k})$ and $\mathcal{F}_{xy}^{2q-\nu+1}(\boldsymbol{k})$.
Using the  relations $\mathcal{F}_{xy}^{\nu}(\boldsymbol{k})=\mathcal{F}_{xy}^{2q-\nu+1}(\boldsymbol{k})$ and $ \mathcal{F}_{xy}^{\nu+1}(\boldsymbol{k})=\mathcal{F}_{xy}^{2q-\nu}(\boldsymbol{k}),$ one can get
\begin{equation}\label{eq:glued_band_constraint}
    C_{+}=C_{-}.
\end{equation}

Applying  Eq.(\ref{eq:sum_rule}) and Eq.(\ref{eq:glued_band_constraint}), we can obtain a  constraint on the Chern number of the glued Dirac band pair analogous to Eq.(\ref{Chern_constraint_2}) of the single band.

\begin{table}[h]
\caption{  \justifying
Chern number sequences of the Hexagonal Lattice Hofstadter Model with different magnetic fluxes $\phi=2\pi p/q \neq \pi$ and hopping coefficients $(t_1,t_2,t_3)$. Each Chern number sequence goes from the lowest to highest bands. The Chern number with label (Dirac) is for glued band pair with Dirac point touchings. For the listed cases, we only observe accidental Dirac point touchings at $E\neq 0$ for $p/q=1/5$ and $(t_1, t_2, t_3)=(1, 2, 3)$.}\label{Table_hex}
\begin{tabular}{|c|c|c|}
\hline
$p/q$ & $(t_1,t_2,t_3)$ & Chern Number  \\
\hline
$1/3$  & $(1, 1, 1)$ & $1, -2, 2(\text{Dirac}), -2, 1$ \\
\hline
$1/4$  & $(1, 1, 1)$ & $1, 1, -3, 2(\text{\text{Dirac}}), -3, 1, 1$ \\ 
\hline  
$1/5$  & $(1, 1, 1)$ & $1, 1, 1, -4, 2 (\text{\text{Dirac}}), -4, 1,1,1$\\
\hline  
$1/3$  & $(1, 2, 3)$ & $1, -2, 1, 1, -2, 1$ \\
\hline 
$1/4$  & $(1, 2, 3)$ & $1, 1, -3, 1, 1, -3, 1, 1$ \\
\hline
$1/5$  & $(1, 2, 3)$ & $1,1,-3 (\text{\text{Dirac}}),1,1,-3 (\text{\text{Dirac}}),1,1$ \\
\hline
$1/3$ & $(6,7,8)$ & $1,-2,2(\text{\text{Dirac}}),-2,1$\\
\hline 
$1/4$ & $(8,9,10)$& $1, 1, -3, 2(\text{\text{Dirac}}), -3,1,1$ \\
\hline
$1/5$ & $(10, 11, 12)$ & $1,1,1,-4,2(\text{\text{Dirac}}), -4,1,1,1$\\ 
\hline
$2/3$ & $(1,1,1)$ & $-1,2,-2(\text{\text{Dirac}}),2,-1$\\
\hline
$2/5$ & $(1,1,1)$ & $3,-2,-2,3,-4(\text{\text{Dirac}}),3,-2,-2,3$\\
\hline  
$3/4$ & $(1,1,1)$ & $-1, -1,3, -2(\text{\text{Dirac}}), 3,-1,-1$\\
\hline
\end{tabular}
\end{table}

     For the symmetric glued band pairs with touching points at $E \neq 0$, $C_+ = C_- \equiv \mathbb{C}'_2$, and the sum rule becomes 
     \begin{equation}
     2\mathbb{C}'_2 + 2\sum\limits^q_{i=1, i\neq \nu,\nu+1} C_i = 0.
     \end{equation}
 With the single-band constraints $pC_i = 1 \mod q$, we get
    \begin{equation}\label{eq:constraint_glued_2}
    -p\mathbb{C}'_2 \equiv (q-2) \mod q \Rightarrow p\mathbb{C}'_2 = 2 \mod q
    \end{equation}
for the glued band pair with touchings at $E\neq 0$.

For the glued  band pair with touchings at $E=0$,  its Chern number $\mathbb{C}_2$ satisfies the sum rule
    \begin{equation}
    \mathbb{C}_2 + 2\sum\limits_{i=1}^{q-1} C_i = 0.
    \end{equation}
 Applying the single-band constraints $pC_i \equiv 1 \mod q$, we get: 
    \begin{equation}\label{eq:constraint_glued_1}
    -\frac{p \mathbb{C}_2}{2} = (q-1) \mod q \Rightarrow p \mathbb{C}_2 = 2 \mod 2q.
    \end{equation}

We numerically calculated the Chern numbers of the glued band pairs at both $E=0$ and $E\neq 0$ at different hopping parameters, and the results verified the constraints in Eq. (\ref{eq:constraint_glued_2}) and (\ref{eq:constraint_glued_1})  for the two cases, as shown in Table \ref{Table_hex}\cite{Dataset}.

\section{Comparison with the topological constraints on the electronic bands in the square lattice}\label{sec:Compare}
In this section, we compare the topological constraints on the electronic bands of the Hofstadter model in the hexagonal lattice  with those in the square lattice. The results are shown in Table \ref{Table_comparison}.

The Hofstadter model in the square lattice with rational flux has been studied extensively in previous works \cite{Square-zero-mode,bernevig2013topological,WEN1989641}. For the reason, we will simply cite some of these results to compare with the results we obtained for the hexagonal lattice.

There are $q$ electronic bands in the square lattice with flux $\phi=2\pi p/q$. For this system, we still choose a $1\times q$ magnetic unit cell.  When $q$ is even, there is an explicit sublattice symmetry of the Hamiltonian\cite{WEN1989641}. This sublattice symmetry anti-commutes with the translation symmetry in the square lattice. For the reason, the energy spectrum of the square lattice Hofstadter model satisfies  
\begin{equation}\label{eq:p-h_symmetry}
E_n(k_x, k_y)=-E_{\bar{n}}(k_x+\pi, k_y)
\end{equation}
when $q$ is even.

Similar to the hexagonal lattice discussed in the last section, if there is a zero energy state at $\mathbf{k}=(k_1, k_2)$ in the system,
the translation symmetry $\mathbf{T}_2$ generates $q$ zero energy states at distinct momenta $\mathbf{k}_j=(k_1+2\pi j p/q, k_2)\ {\rm mod} (2\pi, 2\pi), j=0, 1,..., q-1.$ 
For $q$ even, at $j=q/2$, one gets a zero energy state at $(k_1+\pi, k_2)$ by translation. This state is degenerate with the zero energy state generated by the sublattice symmetry Eq.(\ref{eq:p-h_symmetry}). Each zero energy state then corresponds to a Dirac point at $E=0$ when $q$ is even. 
For $q$ odd, one can not shift the zero energy state at $(k_1, k_2)$ to $(k_1+\pi, k_2)$ by translation.
For the reason, the electronic bands have states crossing the zero energy, but there are no Dirac points at zero energy for $q$ odd \cite{WEN1989641,bernevig2013topological}. 

For square lattice, the zero energy solutions of the Hofstadter Hamiltonian always exist 
regardless of the hopping parameters $t_1, t_2$ \cite{Square-zero-mode,bernevig2013topological}. Moreover, when $q$ is even, the $q$ zero energy states generated by translation are already symmetric with respect to the origin $\mathbf{k}=(0, 0)$, so the ${\cal C}_2$ symmetry does not produce extra zero energy Dirac points \cite{bernevig2013topological}. There are altogether $q$ Dirac points for the square lattice Hofstadter model with $q$ even.

Different from the hexagonal lattice, there are no symmetry enforced Dirac points at $E\neq 0$ in the square lattice Hofstadter model. Moreover, it has been known that the Chern number of the electronic bands in this system can be determined solely by the 
Diophantine equation independent of the hopping parameters \cite{Zak,TKNN}. For the reason, there is no hopping-parameter-tuned topological phase transition in this system, and the accidental Dirac band touchings at $E\neq 0$ do not occur either.

Regardless of whether $q$ even or odd, it has been shown that the Chern number of the square lattice Hofstadter model satisfies \cite{aidelsburger2015artificial}
\begin{equation}\label{CC_square}
C_\mu=C_{q-\mu+1}
\end{equation}
for the  $\mu$-th single band.

Combining Eq.(\ref{Chern_constraint_2}), (\ref{eq:sum_rule}) and (\ref{CC_square}), 
we get the Chern number $\mathbb{C}_2$ of the glued band pair with Dirac point touchings at $E=0$ in the square lattice satisfying
\begin{equation}\label{CC_square_2}
p\mathbb{C}_2=(2-q) {\ \rm mod\ } 2q.
\end{equation}
This is significantly different from the constraint on the glued band with band touchings at $E=0$ in the hexagonal lattice. 

\begin{table}[h]
\centering
\caption{  \justifying
Chern number sequences of the square Lattice Hofstadter Model with different magnetic fluxes $\phi=2\pi p/q \neq \pi$. Each Chern number sequence goes from the lowest to highest bands. The Chern number with label (Dirac) is for glued band pair with Dirac point touchings. The Chern number in this system is independent of the hopping parameters.}\label{Table_square}
 \begin{tabular}{|c|c|}
\hline
$p/q$ & Chern Number  \\
\hline
\ \ \ $1/3$ \ \ \ & $1,-2,1$ \\
\hline
\ \ \  $1/4$ \ \ \ & $1,-2(\text{Dirac}),1$ \\
\hline
\ \ \  $1/5$ \ \  \ & $1,1,-4,1,1$ \\
\hline
\ \ \ $1/6$ \ \ \  & $1,1,-4(\text{Dirac}),1,1$ \\
\hline  
\ \ \ $2/3$ \ \ \  & $-1,2,-1$\\
\hline
\ \ \ $2/5$ \ \ \  & $-2,3,-2,3,-2$\\
\hline
\ \ \ $3/4$ \ \ \ & $-1,2(\text{Dirac}),-1$\\
\hline
\ \ \  $3/5$ \ \ \ & $2,-3,2,-3,2$\\
\hline 
\ \ \ $5/6$ \ \ \ & \ $-1,-1,4(\text{Dirac}),-1,-1$\ \ \\
\hline 
\end{tabular}
\end{table}

We have computed the Chern number of the square lattice Hofstadter model with different flux numerically, as shown in Table \ref{Table_square}\cite{Dataset}, and verified that they satisfy the constraint in Eq.(\ref{Chern_constraint_2}), (\ref{CC_square}) and (\ref{CC_square_2}). 

At last, the implicit sublattice symmetry in the square lattice Hofstadter model at $q$ even anti-commutes with the translation symmetry of the primitice lattice. For this reason, when the primitive translation symmetry  in one direction is broken, the Dirac band touchings at zero energy disappear and one can get a topological insulator with a Mobius twist in the edge band \cite{Z2,Xiao2024,Dimerized-Hofstadter-Model},
whether the Chern number of the system is zero or not. This is also different from the consequence of the translation symmetry breaking in the hexagonal lattice as discussed in Sec.\ref{sec:Chern}.A.

\section{Discussions and Summary}\label{sec:summary}
Real materials usually do not have the ideal hexagonal lattice structure \cite{Dean2013}. 
However, our derivations in the previous sections show that the results we obtained  remain valid as long as  the translation and sublattice symmetry of the hexagonal lattice are preserved. For the reason, the results in this work are applicable to most real materials with moderate distortions.

The breaking of the translation symmetry corresponds to the breaking of the periodicity of the primitive lattice, e.g., when charge or spin density waves appear in the system or adding stagered potential on the $AA$ or $BB$ sublattice in the MUC. 
With the breaking of $\mathbf{T}_2$ symmetry, the constraints on the Chern number Eq.(\ref{eq:Chern_constraint_1}),(\ref{Chern_constraint_2}), (\ref{eq:constraint_glued_2}) and (\ref{eq:constraint_glued_1})
will no longer be valid. The breaking of $\mathbf{T}_2$ will lift the band touchings at $E\neq 0$ for the $\phi=\pi$ case in the hexagonal lattice, but it does not necessarily remove the Dirac points at $E=0$. To lift the Dirac band touchings at $E=0$, the sublattice symmetry needs to be broken in the hexagonal lattice. To do this, one way is to add diagonal terms in the Hamiltonian matrix in Eq.(\ref{eq:Hamiltonian_k}), e.g., by adding potentials on the $AB$ sublattice or introducing next nearest neighbor hopping. Another way is to break the off-diagonal symmetry of the Hamiltonian, e.g., by staggerizing the hopping between the $AB$ sublattice. The effects of breaking the sublattice and translation symmetries have been shown in Fig.\ref{fig:sublattice_broken} and Fig.\ref{fig:Translation_Broken}.

The hexagonal Hofstadter model can be experimentally  realized in superlattices with nanometer scale unit cells, such as Moire bilayer graphene\cite{Dean2013}, or graphene on a properly aligned Boron-Nitride substrate\cite{Geim2013, Hunt2013, Spanton2018}.  In these systems, a suitable commensurate magnetic flux can be reached in a super unit cell under experimentally feasible magnetic field. The Hofstadter model can also be simulated by cold atoms in optical lattices shown in Ref.\cite{Bloch2013, Ketterle2013}, or photonic crystals in Ref.\cite{Kraus2012, Lahini2009}. Particularly, a recent experiment realized a $\pi$ flux Hofstadter model in square lattice by controlling the higher-orbital degrees of freedom in confined-Mie-resonance photonic crystals~\cite{Huang2025}. 
These systems provide a variety of promising platforms to verify the theoretical results of the hexagonal Hofstadter model in this work.

In summary, we analyzed how the projective lattice symmetry shapes the electronic band structure of a hexagonal lattice in a magnetic field. We demonstrated that at flux $\phi=\pi$, the projective lattice symmetry robustly enforces Dirac points at energies $E\neq 0$, independent of the hopping parameters. By contrast, the presence of Dirac points at $E=0$ depends jointly on the hopping parameters and the lattice symmetry. When the hopping parameters admit zero-energy solutions of the Hamiltonian, the sublattice, magnetic translation, and ${\cal C}_2$ rotation symmetry together generate $2q$ Dirac points at $E=0$ in general case. By tuning the hopping parameters, the $2q$ Dirac points  merge to $q$ semi-Dirac points at the phase transition points from gapless states to gapped states. We also examined the topological constraints that the lattice symmetry imposes on the Chern numbers of electronic bands in the hexagonal lattice, addressing both isolated gapped bands and symmetry-enforced band multiplets connected by Dirac-point touchings.
Finally, we compared these results for the hexagonal lattice with their counterparts in the square lattice, and demonstrated the key similarities and differences between the two systems.

\section{Acknowledgements}
We thank the helpful discussions with Y. X. Zhao and Rong Xiao. This work is supported by the NNSF of China under Grant No. 11974166 and the NSF of Jiangsu Province under Grant No.BK20231398. 

\appendix

\begin{widetext}

\section{Band expansion around the degenerate  point at $E\neq 0$}
In this appendix, we expand the electronic band near the degenerate point at $E\neq 0$ for the flux $\phi=\pi$ case and show that the dispersion is linear around this point. 

The eigenvalue of the tight-binding Hamiltonian for the $\pi$ flux case can be solved from Eq.(\ref{eq:TBG_pi_flux}) and we get 
\begin{align}\label{eigenenergy}
+E_{1}(k_1,k_2)&=\sqrt{t_1^2+t_2^2+t_3^2-\sqrt{2}\sqrt{t_1^2t_2^2+(t_1^2+t_2^2)t_3^2+t_1^2t_3^2\cos(2k_1)+t_2^2(t_1^2\cos(2k_1-k_2)-t_3^2\cos(k_2))}},\notag\\ 
+E_{2}(k_1,k_2)&=\sqrt{t_1^2+t_2^2+t_3^2+\sqrt{2}\sqrt{t_1^2t_2^2+(t_1^2+t_2^2)t_3^2+t_1^2t_3^2\cos(2k_1)+t_2^2(t_1^2\cos(2k_1-k_2)-t_3^2\cos(k_2))}},\notag\\ 
-E_{1}(k_1,k_2)&=-\sqrt{t_1^2+t_2^2+t_3^2-\sqrt{2}\sqrt{t_1^2t_2^2+(t_1^2+t_2^2)t_3^2+t_1^2t_3^2\cos(2k_1)+t_2^2(t_1^2\cos(2k_1-k_2)-t_3^2\cos(k_2))}},\notag\\ 
-E_{2}(k_1,k_2)&=-\sqrt{t_1^2+t_2^2+t_3^2+\sqrt{2}\sqrt{t_1^2t_2^2+(t_1^2+t_2^2)t_3^2+t_1^2t_3^2\cos(2k_1)+t_2^2(t_1^2\cos(2k_1-k_2)-t_3^2\cos(k_2))}}.
\end{align}

It can be verified that $E_{1}(\frac{\pi}{2},0)=E_{2}(\frac{\pi}{2},0),$ which correspond to the finite energy degenerate points at $\boldsymbol{X}=(\frac{\pi}{2},0).$ We expand $E_1(k_1,k_2)$ near the degenerate point at  $\boldsymbol{X}$ and get   
 \begin{equation}
     E_{1}(k_1,k_2)\approx\sqrt{t_1^2+t_2^2+t_3^2}-\frac{\sqrt{2}}{2\sqrt{t_1^2+t_2^2+t_3^2}}\eta(k_1,k_2),
 \end{equation}
where 
\begin{equation}
\eta(k_1,k_2)\approx\sqrt{2t_1^2(t_2^2+t_3^2)(k_1-\frac{\pi}{2})^{2}-2t_1^2t_2^2(k_1-\frac{\pi}{2})k_2+\frac{1}{2}t_2^2(t_1^2+t_3^2)k_2^2}.
\end{equation}
The energy dispersion near the degenerate point  $X$ is then linear to the momentum $(k_1-\pi/2, k_2)$ in all directions and $X$ is a Dirac point.

\section{Band expansion around the zero-energy degenerate points} 

In this appendix, we expand the electronic bands around the degenerate points at $E=0$, including the case with $2q$ degenerate points and the case when the $2q$ degenerate points merge to $q$ semi-Dirac points at the critical states. 

We first expand the dispersion around the $E=0$ degenerate points for the isotropic case with $t_1=t_2=t_3=1$ and $\phi=\pi$ shown in Fig.\ref{fig:3D Band}. There are $4$ degenerate points at $E=0$ in this case. We expand the energy $E_1(k_1, k_2)$ in Eq.(\ref{eigenenergy})  around the degenerate point at momentum $(k_1,k_2)=(\frac{5\pi}{6},-\frac{2\pi}{3})$ and get 
\begin{equation}
E_{1}(k_1,k_2)=\frac{1}{\sqrt{6}}\sqrt{
[2(k_1-\frac{5\pi}{6})-\frac{1}{2}(k_2+\frac{2\pi}{3})]^2+\frac{3}{4}(k_2+\frac{2\pi}{3})^2}.
\end{equation}
The dispersion is then linear in all directions near the degenerate point and it is a Dirac point.

For the critical states of $q$ degenerate points,  we consider the case shown in  Fig.\ref{fig:q_Dirac_points}(a) with $t_1=\sqrt{2},t_2=t_3=1$ and $\phi=\pi$. A band touching point exists at the momentum $(0,0)$ in this case as $\pm E_1(0, 0)=0$.  
We  perform an expansion of $E_{1}(k_1,k_2)$ in Eq.(\ref{eigenenergy}) around $(0,0)$ and get 
\begin{equation}
E_1(k_1,k_2)\approx \sqrt{-k_1k_2+2k_1^2+\frac{1}{8}k_2^2}=\vert \sqrt{2}k_1-\frac{1}{2\sqrt{2}}k_2\vert.
\end{equation}
We can see that along the direction $k_2=4k_1$, the linear dispersion vanishes and the lowest order dispersion is parobolic. In the other directions, the dispersion  is still linear to the momentum $(k_1, k_2)$. 
The degenerate points at $E=0$ at this phase transition point are then semi-Dirac points.

\end{widetext}

\clearpage

\end{document}